\documentclass[11pt, titlepage]{article}
\usepackage{eurosym}
\usepackage{amsmath,amssymb,amsthm,amsfonts}
\usepackage[mathscr]{eucal}
\usepackage{graphicx}
\usepackage{epstopdf}

\usepackage{lineno,hyperref}
\modulolinenumbers[5]

\begin{document}


\author{Cristina Coscia\thanks{
Department of Architecture and Design, Politecnico di Torino. } Roberto Fontana \thanks{
Department of Mathematical Sciences, Politecnico di Torino} and Patrizia
Semeraro\thanks{
Corresponding author.  Department of Architecture and Design, Politecnico di Torino.
Viale Mattioli 39, 10125 Torino. e-mail address: patrizia.semeraro@polito.it} }

\title{Graphical models for studying  museum networks:\\
 the Abbonamento Musei Torino Piemonte}

%
%
\date{}
\maketitle

%
%
%
%

\begin{abstract}

Probabilistic  graphical models are a powerful tool to represent real-word phenomena and to learn network structures starting from data. This paper applies graphical models in a new framework to study association rules driven by consumer choices in a network of museums. The network consists of the museums participating in the program of  Abbonamento Musei Torino Piemonte, which is a yearly subscription  managed by the {\it Associazione Torino Citt\`a Capitale Europea}. Consumers are card-holders, who are allowed to entry to all the museums in the network for one year. We employ graphical models to highlight associations among the museums driven by card-holder visiting behaviour. We use both undirected graphs to investigate the strength of the  network and directed graphs to highlight asimmetry in the association rules.



\end{abstract}
\titlepage

\section{Introduction}

Probabilistic graphical models are a powerful tool to represent real-word phenomena and to learn network structures starting from data.  In cultural goods frameworks the structure of data is usually used to define  clusters of  consumers with the aim of supporting investment strategies, e.g. pricing policies. In this paper we study the network of museums participating to the project of  Abbonamento Musei Torino Piemonte (AMTP network). AMTP is a yearly subscription  managed by the {\it Associazione Torino Citt\`a Capitale Europea} (ATCCE)\footnote{\url{www.abbonamentomusei.it/pages/Abbonamento_Musei}}.  The 2007 AMTP database was analyzsed in
 \cite{BrondC} with the  goal of clustering subscribers  according to their visiting behaviour. Here we propose  a different approach.
Instead of clustering consumers in the network, we  investigate museums association driven by consumer behaviour, with the aim of analyzing the strength of the network.
In this paper we apply graphical models in a new framework to study conditional independence structure of the museums belonging to the AMTP network.
This network was created in 1995 and is  available to people living in the Piemonte region (Italy).   
For a yearly subscription fee, AMTP card-holders have  free entry to all the museums and  all the temporary/permanent exhibitions participating in the program for the subscription year, from January to December. 
 In recent years the number of subscribers has increased enormously: from 3,279 cards in 1999 to 82,802 cards in 2012.

 In this work we  analyze the structure of the  2012  AMTP network.  We focus on the conditional independence structure  induced by subscribers' visiting behaviour. A descriptive analysis of the association structure of the 2012 AMTP network was conducted  in \cite{cosciamarket};  the authors  performed a market basket analysis, assuming that  consumers are card-holders,  products are museums participating in the program and  the unit of shopping is one year.

The aim of our analysis  is to  learn  the structure of the joint probability distribution of the 2012 AMTP network from data. Indeed, we  use graphical models to perform structural
learning from the AMTP database. On  the basis of the observed data, i.e. the visiting sequence or itinerary of each card-holder in 2012, we want to choose the graph
whose independency model best represents the mechanism induced by the consumers' choice to visit a set of museums. Consumers are classified according to the presence of previous subscriptions to the AMTP to point out possible differences in the visiting behaviour and consequent influences on the network of museums.
 We consider both undirected and directed graphical models.

 Undirected graphical models encode the conditional independence structure of the joint distribution of the  network. The conditional independence structure of the network highlights possible independent sub-networks of museums and provides a useful indication about the strength of the network. Given the high-dimensionality of our database, we consider the simplest form of undirected graphical models, i.e.  forest and tree graphs. The connected components (trees) of the forest represent independent sub-networks of museums. The presence of a low number of connected components would indicate that museums in  the AMTP network are connected by paths which represent the consumers' choice to associate museums in their annual visiting itinerary. 

 Directed graphical models allow us to highlight asymmetry in the network association structure. We focus on   directed acyclic graphs (DAG), where asimmetry comes from the causal mechanism between variables in the network. In this framework, DAG  models imply an ordering among the museums, which defines a statistical time, as defined in \cite{pearltheory}.  Conjecture 1 in  \cite{pearltheory} states a link between physical time and statistical time. According to this conjecture, we investigate whether the mechanism underlining the causal ordering among museums is driven by the visiting itinerary of card-holders. We look for a correspondence between the physical time induced by the visiting itinerary, and the statistical time as determined by the DAG  learned from data.  

The paper is organized as follows. Section \ref{Data} introduces the data, Section \ref{Meth} presents the methodology and recalls both directed and undirected graphical models. Section \ref{Appl} discusses the results and Section \ref{Concl} concludes.

\section{Data}\label{Data}

 In this paper we analyze the 
2012 AMTP transaction database.  The 2012 AMTP database collects information for each card-holder about the museums visited, the time of each visit and the presence of previous subscriptions. The number of card-holders is  82,602. The museums participating in the program in 2012
are $158$, including $15$
temporary exhibitions and 143
permanent museums. Our analysis focus on the 23  main museums, i.e museums with a number of visits higher that $85$th percentile of the number of  visits (that is around 4,000 visits). 
Table \ref{CM} maps the 23 museums with their codes.
    
\begin{center}
Insert Table  \ref{CM} here
\end{center}

We analyse association between the main museums driven by consumer visiting behaviour. We consider which museum has been visited by each card-holder, regardless the number of times she returned in the same museum. Therefore we do not  consider  repeated visits and we have a total   $287,259$ visits.
The most visited museum in 2012 is the {\it Reggia di Venaria Reale}, with 21.3\% of total visits. Descriptive statistics for this databsase can be found in \cite{cosciamarket}.

\section{Methodology}\label{Meth}

Every  museum participating in the program in year 2012  is modelled by a variable  $M_i$, so that $M_i=1$  is the event that museum $i$ is visited by a card holder and $M_i=0$ its complement.   We also have a variable $S$ to classify card-holders, $S$ can take three levels: level $0$ if  card-holder has the subscription for the first time in 2012 (new card-holder), level $1$ if the card-holder renewed the subscription of the past year (renewal card holder) and level $2$ if the  card-holder has been a subscriber, but not in the past year (old card-holder).  
Museums association driven by consumer behaviour is then represented by association among the variables $M_i$. The count $n_i$ is the number of times the museum $M_i$ has been visted in one year.

We recall here the graphical models we use to study museums associations driven by consumer behaviour. 
A graph $G$ is  described by the pair $(V, E)$, where $V$ is a set of vertices, each corresponding
to a random variable, and $E$ is the set of edges between such vertices. In our framework statistical variables are museums $M_i$ or museums plus card-holder classification $S$, which we identify with the vertices of a graph, i.e. $V=\{M_i, i=1,\dots, n\}$, $n=23$ or $V_S=V\cup \{S\}$. 
 Graphical models are classes of multivariate distributions whose conditional independence properties are encoded by a graph. Graphs may be directed or undirected (see \cite{koller2009probabilistic} as standard reference for graphical models).

The aim of our analysis  is structural
learning from the AMTP database, which means to learn  the structure of the joint probability distribution of museums from data. On  the basis of the observed data, we want to choose the graph
whose independency model best represents the mechanism induced by the consumers' choice to visit a set of museums in $V$. 
We shall
consider here both undirected edges, such that, if $(M_i, M_j) \in  E$ also $(M_j, M_i)\in  E $ and directed
edges, for which only one of the previous holds. We name {\it arc} a directed edge.
For any triplet of disjoint subsets $A, B, S \subset V$ such that $A, B$ are non-empty, one may evaluate
that A is conditional independent from $B$ given $S$ in a graph $g = (V, E)$.

In the case of undirected graphs,  vertices are connected by an edge when the corresponding variables are not conditionally independent given the other variables in the graph. 
Formally, if $V=\{M_i, i=1,\dots, n\}$ is a finite set of vertices, the set of edges is a subset $\{(M_i, M_j), i<j\}$ of $V\times V$. The absence of an edge $e_{ij}=(M_i, M_j)$ means that $M_i$ is conditionaly independent by $M_j$ given all the remaining variables, formally:
\begin{equation}\label{Cind}
M_i\bot M_j|(V\setminus\{M_i, M_j\}).
\end{equation}
Conditon \ref{Cind} is the pairwise Markov property. Hence a graph is a model of conditional independence. 

We would like to compare all possible
graphs for our given set of museums. The high-dimensionality of our database provides a challenging scenario, since we have $2^{23(23-1)/2}=1.44*10^{76}$ possible undirected graphs. So we confine our analysis to forest and tree graphs. A forest is an undirected graph with no cycles, which may be composed of several connected components, called trees.  We measure the importance of museums in the network using three network metrics: degree, betweenness  centrality and closeness centrality. The degree of the node in the number of edges incident on it. The betweenness  centrality  is a measure of the degree to which a given node lies on the shortest paths (geodetics) between other nodes in the graph and the closeness centrality rates the centrality of a node by its closeness (distance) to other nodes. We computed the measures according to \cite{butts2010sna}.

The notion of directed graphs is less intuitive and we recommend the interested
reader to consult \cite{lauritzen1990independence}  or \cite{pearl1987logic}.
Here we consider  directed
acyclic graphical models (DAG).
DAG models are determined by directed graphs that do not contain directed cycles.
We have only  discrete variables, so we shall concentrate on graphs
representing them. 
The DAG defines a factorization of the joint probability distribution of $V$ into a set of local distributions, one for each variable. The form of the factorization is given by the Markov property of Bayesian networks which state that every random variable $M_j$ directly depends only on a set of vertices $\Pi_{M_j}\subset V\setminus \{M_j\}$, named parents of $M_j$.  Parents of a variable $M_j$ are identified using the conditional independence structure of the joint probability of $V$.  Namely, from the chain rule of probability, we have:

\begin{equation}
P(M_1,\dots, M_n)=\prod_{i=1}^nP(M_i|\{ M_{i-1}, \dots, M_1\}),
\end{equation}

for each $M_i$. Let $\Pi_{M_i}\subset \{ M_{i-1}, \dots, M_1\}$ be a set of nodes and suppose that  the variables $M_i$ and $\{ M_{i-1}, \dots, M_1\}$ are  conditionally independent given $\Pi_{M_i}$, i.e. 

\begin{equation}
P(M_i|\{ M_{i-1}, \dots, M_1\})=P(M_i|\Pi_{M_i}).
\end{equation}
Hence, the Markov property of Bayesian network is:
\begin{equation}
P(M_1,\dots, M_n)=\prod_{i=1}^nP(M_i|\Pi_{ M_j}).
\end{equation}

The arcs resulting from the estimated DAG model describe a causal relationship between each variable and its parents. Any ordering of the variables that agrees with the DAG structure estimated is called statistical time (\cite{pearltheory}). Conjecture 1 in \cite{pearltheory} asserts that in most natural phenomena the physical time coincides with at least one statistical time. In our framework, physical time induces an ordering between any pair of museums according to the number of times that one of them has been visited before. 
We empirically verify Conjecture 1, by comparing the statistical order arising from DAG, with the order induced by physical time.

\subsection{Computational aspects}
We perform the analysis using a standard laptop (CPU Intel core I7-2620M CPU 2.70GHz 2.70GHz, RAM 8GB). We used SAS and  R. In particular we used the packages gRaphHD \cite{de2009high}, Bnlearn \cite{scutari2009learning} and SNA \cite{butts2010sna}.

\section{Application and discussion of results}\label{Appl}

As we already described we focus on the main $23$ museums. We have $23$ binary variables $M_i, i=1,\dots,23$ representing the $23$ museums. The variable are codified to take the value $M_ i=1$ if card-holder visited  the museum $M_i$ and  to take the value $M_i=0$  otherwise.  We also have the three levels variable $S$ to classify card-holders.  

\begin{figure}[htbp]
\centering
\includegraphics[width=0.7\columnwidth]{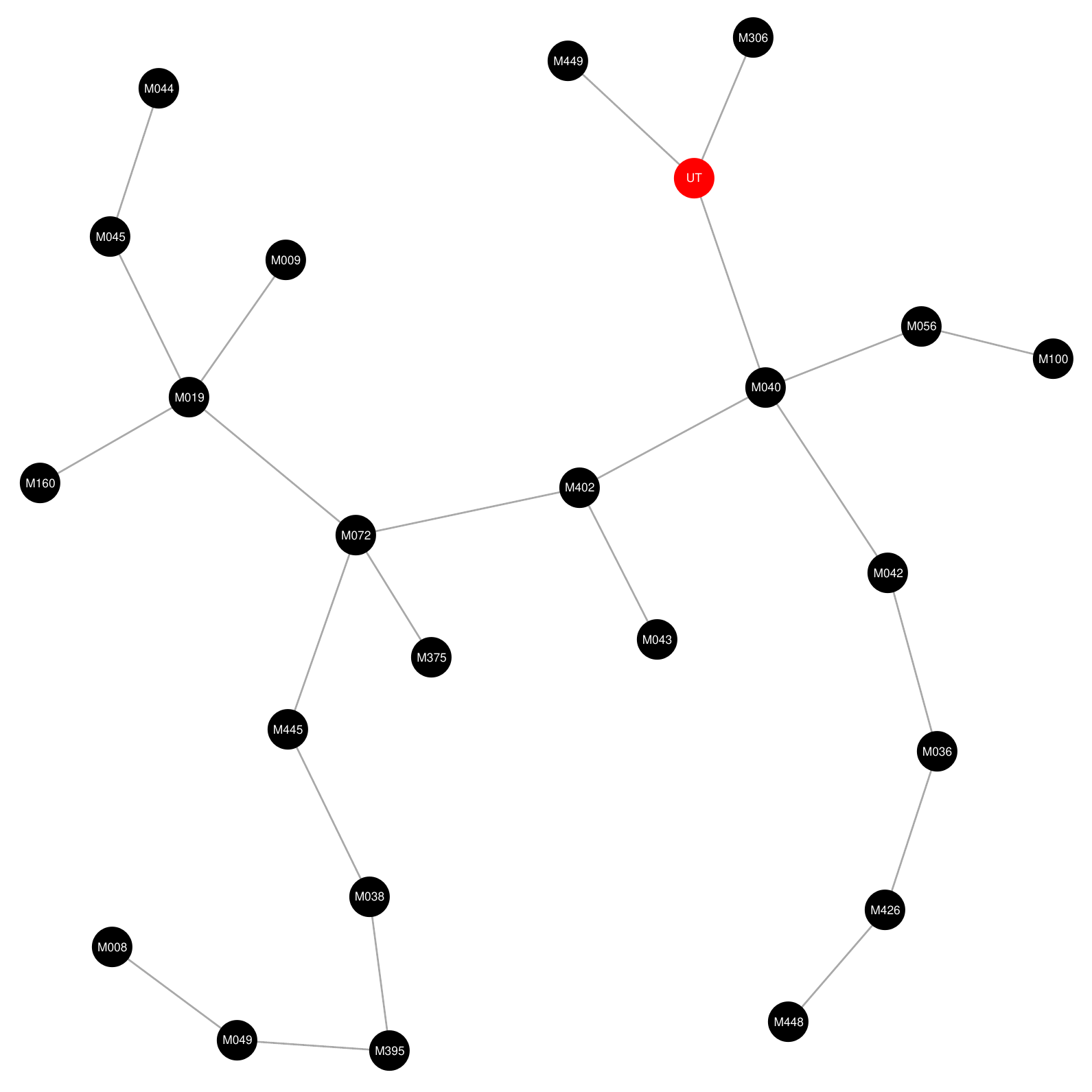}
\caption{Minimum BIC forest, set of nodes: $\mathcal{M}\cup\{T_H\}$.}
\label{T_red}
\end{figure}

 Figure \ref{T_red} shows the conditional independence structure of  museums visits and type of card-holder through a minimum BIC forest; the  node UT  is the variable $S$.

\begin{center}
Insert Figure \ref{T_red} here
\end{center}

\begin{figure}[htbp]
\centering
\includegraphics[width=0.7\columnwidth]{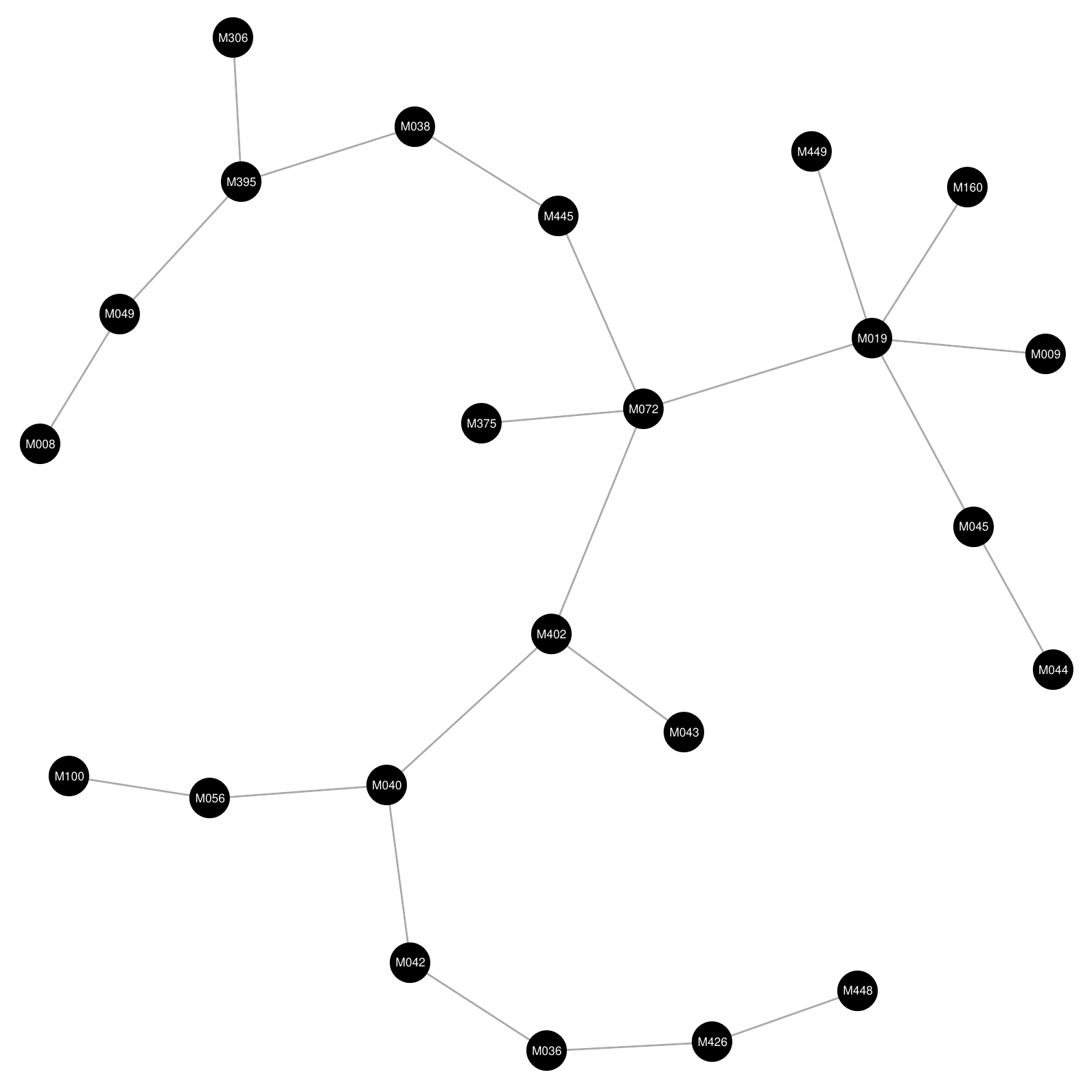}
\caption{Minimum BIC forest, set of nodes: $\mathcal{M}$.}
\label{T_red_nout}
\end{figure}

Figure  \ref{T_red_nout} shows the conditional independence structure of  museums visits without the variable $S$ through a minimum BIC forest.

\begin{center}
Insert Figure \ref{T_red_nout} here
\end{center}

 Table \ref{degree} shows degree,   betweeness and closeness of each nodes.

\begin{center}
Insert Table \ref{degree} here
\end{center}

The node $S$ has degree three, and both betwennes and closeness are not too low. Novertheless, comparing the two figures, we immediately observe that the variable $S$ is not a key variable for the dependence structure induced by consumers.    The only associations influenced by considering if consumers are new subscribers are the temporary exhibition {\it A un passo da Degas} and the museum {\it Forte Bard}. The temporary exhibition was a very important cultural event for the city of Turin and it was strongly advertized, while the Forte Bard organized in 2012 several cultural events and temporary exhibitions, as the Alberto Giacometti exhibition.  Thus, association structure induced by visiting behaviour seems to be independent from the state of the consumer except for these two cases concerning new important cultural events. For this reason we discuss together the two graphs, highlighting the few differences due to the presence or absence of the variable $S$.
First of all we observe that we have only one tree, thus all museums are connected. This result underlines the strength of the network, since no museum stands alone and is independent from the remaining part of the structure.  
We have three central nodes: {\it Museo Egizio} (M040), {\it Palazzo Madama} (M072) and {\it Palazzo Reale} (M402), which exibits the highest levels of betwennes and closeness (see Table \ref{degree}). These museums are hystorical museums in the center of the city and define a traditional consumption bundle of museums enforced by their spatial proximity.
 In particular {\it Museo Egizio} is very important for the city of Turin, being considered the second Egyptian museum in the world after Cairo  museum.  
For both the graphs the central node is {\it Palazzo Reale}, which divides the graph into two connected components, one including {\it Museo Egizio} and the other including {\it  Palazzo Madama}. The two connected components have serveral nodes and can be considered as two subnetworks linked by {\it Palazzo reale}.  
Another museum very important for the city of Turin is the {\it Cinema} museum (M042), also in the center of the city and located in the {\it Mole Antonelliana}, the  tallest museum in the world.  The {\it Mole Antonelliana}  is a major landmark building in Turin and  is named after the architect who built it, Alessandro Antonelli. 
Nevertheless, in this network its role is less central, and it results directly connected only with {\it Museo Egizio} and {\it Museo dell'automobile} (M036), another important museum for Turin. Together with {\it Museo Egizio} and {\it Palazzo Madama} the node {GAM} (M019) has the highest degree. Indeed, it is one of the main contemporary art museum of the city. It is directly connected with {\it Rivoli} palace (M009) which, although it is an hystorical castle, is a famous location for temporary exibition and {\it Pinacoteca Agnelli} (M160), which is a private foundation and displays a collection of art works between the XVII and XX century.

\begin{figure}[htbp]
\centering
\includegraphics[width=0.7\columnwidth]{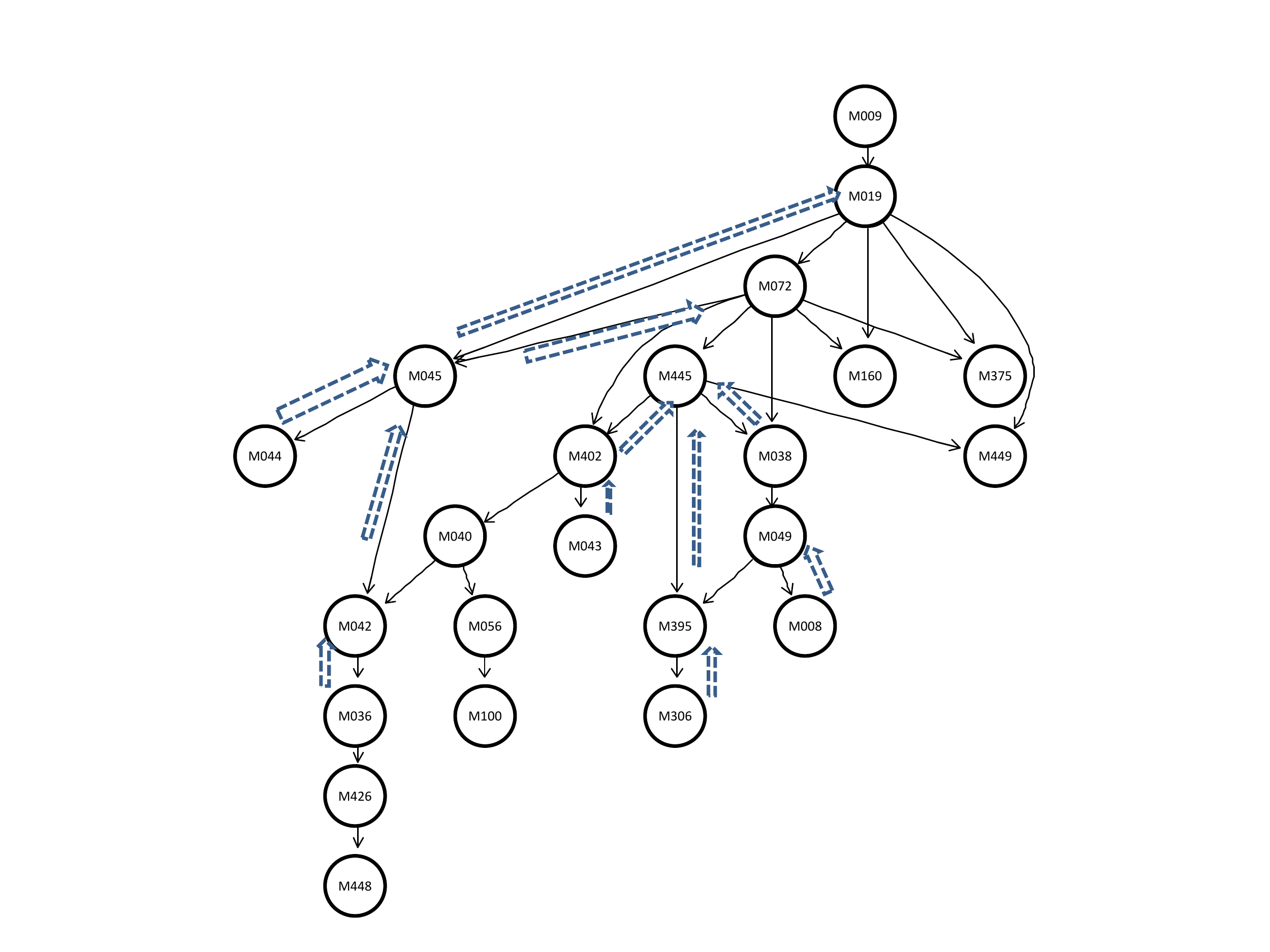}
\caption{DAG graphical model. BIC criterium, penalty 200.}
\label{s_complex}
\end{figure}

Now we switch to the analysis of asimmetry of associations by using DAG models. Since we aim to verify Conjecture 1 comparing statistical time with physical time, where physical time is defined by the visiting itinerary, we do not include the variable $S$ in the set of nodes of the graph. To explain a causal relationship between pair of museums, it looks reasonable to consider which one has been visited before in time. This fact could link statistical time with physical time, through card-holders itineraries.
We start analysing statistical time arising from a DAG graphical model explaining museums' causal relatioships driven by all card-holders' visits.
Figure \ref{s_complex} is the DAG graph without the variable $S$.

\begin{center}
Insert Figure \ref{s_complex} here
\end{center}

 Each arc in the graph indicates a causal relationship between the pair of  nodes, each dashed line between a pair of museums indicates  which one has been visited before in time.  Looking at each pair directly linked by one arc, the visiting time generates an ordering which is represented by the dashed lines in the figure. Comparing arcs and dashed lines we notice that the two ordering are very similar, but with some exceptions, as e.g. the pair {\it Museo Regionale di Scienze Naturali} and {Castello di Rivoli}. This means that although museum  {\it Museo Regionale di Scienze naturali} (M045) has been visited before, its visit has been caused by the (future) visit of museum {\it Castello di Rivoli}, which sounds very strange. Indeed, probably there are some unobservable variables inducing  causal relationships between museums. Actually, one possible variable is $S$, which is observable. We employed the DAG model for the three levels of $S$ and we found different statistical times, although the consumer classification  does not effect the strength of the network, as revealed from the first step of analysis. Since also in these cases statistical time and physical time are sometimes inverted for brevity we do not report the graphs.  
 Nevertheless, the strength of the network is confirmed also by the DAG structure, where again all museums are connceted. Also the central role of   {\it Museo Egizio, Palazzo Madama} and {\it Palazzo Reale} is confirmed, since they have parents and sons and morevore they are directly linked, being {\it Palazzo Madama} a partent of {\it Palazzo Reale}, which is a parent of {\it Museo Egizio}.

\section{Conclusion}\label{Concl}
This paper applies graphical models in a new framework to explore association among museums in the AMTP network driven by consumer visiting behaviour.  We analyze the strength of the association structure with undirected graphical models. Due to the high dimensionality of data we confine the analysis to the simplest form, i.e. trees and forests. We find only one tree connecting all the museums, supporting the strength of the network. The central nodes are the main museums of the city, which are historical museums in the center of the city.

We also use directed graphical models to highlight assimmetries in the association structure of the network. We consider DAG models where asimmetry comes from the causal mechanism between variables. We investigate whether the causal relationship between museums  inferred from data corresponds to  physical time. Physical time is defined by the visiting itineraries of card-holders. We found some discrepancies between statistical time and the visting itinerary, highlighting the possible presence of unobservable factors explaining the causal relationship between museums. 

Further research aims at investigating the influence of personal details on visting behaviour. In fact,  from 2014 ATCCE started to collect also personal information of subscribers.

\noindent {\bf Acknowledgments.}
The authors thank Cinzia Carota, Alessandra Durio and Marco Guerzoni for useful discussions and Francesca Leon for providing the database of AMTP card-holders.


\newpage

\begin{small}
\begin{center}
\begin{table}
\begin{tabular}{|l|l|}
\hline
\label{CM}
 Code & Museum/temporary exibition\\
\hline
M008 & CASTELLO DI RACCONIGI\\
M009 & ASTELLO DI RIVOLI - MUSEO D'ARTE CONTEMPORANEA\\
 M019 & GAM - GALLERIA CIVICA D'ARTE MODERNA E CONTEMPORANEA\\
M036 & MUSEO DELL' AUTOMOBILE CARLO BISCARETTI DI RUFFIA\\
 M038 & MUSEO DI ARTI DECORATIVE FONDAZIONE PIETRO ACCORSI\\
 M040 & MUSEO EGIZIO\\
 M042 &  MUSEO NAZIONALE DEL CINEMA\\
M043 & MUSEO NAZIONALE DEL RISORGIMENTO ITALIANO\\
 M044 & MUSEO NAZIONALE DELLA MONTAGNA DUCA DEGLI ABRUZZI\\
M045 & MUSEO REGIONALE DI SCIENZE NATURALI\\
M049 & PALAZZINA DI CACCIA DI STUPINIGI\\
M056 & REALI TOMBE DI CASA SAVOIA - BASILICA DI SUPERGA\\
 M072 & MUSEO CIVICO D'ARTE ANTICA E PALAZZO MADAMA\\
M100 & ABBAZIA SACRA DI SAN MICHELE\\
M160 & PINACOTECA GIOVANNI E MARELLA AGNELLI\\
M306 & FORTE BARD\\
M375 & MAO MUSEO D'ARTE ORIENTALE\\
 M395 & REGGIA DI VENARIA REALE\\
 M402 & PALAZZO REALE\\
M426 & OGR\\
 M445 & MOSTRA - Torino, Europa - Le Grandi opere d'arte della Galleria Sabauda\\
M448 & MOSTRA VOLARE\\
 M449 & Mostra "A un passo da Degas"\\
\hline
\end{tabular}
\caption{Museum and temporary exibition codes }
\end{table}
\end{center}
\end{small}

\begin{small}
\begin{center}
\begin{table}
\begin{tabular}{|l|r|r|r|}
\hline
\label{degree}
node & degree & betweenness & closeness \\     
\hline                                                                        
UT & 3 & 86 & 0.26 \\
M008 & 1 & 0 & 0.15 \\                                                                              
M009 & 1 & 0 & 0.22 \\                                                                               
M019 & 4 & 162 & 0.28 \\                                                                               
M036 & 2 & 84 & 0.22 \\                                                                               
M038 & 2 & 120 & 0.24  \\                                                                              
M040 & 4 & 304 & 0.32 \\                                                                               
M042 & 2 & 120 & 0.26 \\                                                                               
M043 & 1 & 0 & 0.26 \\                                                                               
M044 & 1 & 0 & 0.19 \\                                                                               
M045 & 2 & 44 & 0.23 \\                                                                               
M049 & 2 & 44 & 0.17 \\                                                                               
M056 & 2 & 44 & 0.25 \\                                                                               
M072 & 4 & 334 & 0.34 \\                                                                               
M100 & 1 & 0 & 0.2  \\                                                                               
M160 & 1 & 0 & 0.22 \\                                                                               
M306 & 1 & 0 & 0.21 \\                                                                               
M375 & 1 & 0 & 0.26 \\                                                                               
M395 & 2 & 84 & 0.2  \\                                                                               
M402 & 3 & 284 & 0.34 \\                                                                               
M426 & 2 & 44 & 0.18   \\                                                                             
M445 & 2 & 152 & 0.28 \\                                                                               
M448 & 1 & 0 & 0.16 \\                                                                               
M449 & 1 & 0 & 0.21\\
\hline
\end{tabular}
\caption{Museum metrics: degree, betweenness and closeness }
\end{table}
\end{center}
\end{small}

\end{document}